\documentclass[letterpaper, 10 pt, conference]{ieeeconf}  
\IEEEoverridecommandlockouts         
\usepackage[utf8]{inputenc}
\usepackage[T1]{fontenc}
\usepackage[american]{babel}
\usepackage[autostyle]{csquotes}
\usepackage{booktabs} % for rulers in tables
\usepackage[final]{microtype} % for \microtypesetup{protrusion=true}
\usepackage[hidelinks, breaklinks = true]{hyperref} % hidelinks removes colored boxes around references and links, breaklinks allows linebreak in e.g. list of figures for long captions
\usepackage{amssymb}

\usepackage{amsthm}
\usepackage{amsmath}
\usepackage{algorithm} %for algorithm environment [noend]
\usepackage{algpseudocode} %for algorithm environment
\usepackage{units} % for units
\usepackage{mathtools}

\usepackage{graphicx}
\usepackage{psfrag}
\usepackage{tikz}
\usepackage{comment}	
\usepackage{acronym}
\usepackage{mathrsfs}
\usepackage{subcaption}
\usepackage{fixltx2e}
\usepackage{amsfonts}
\usepackage[capitalise]{cleveref}
\usepackage{hyperref}

\usepackage{bbm}
\usepackage{scalerel}
\usepackage{algorithm}
\usepackage{algpseudocode}
\usepackage{multirow}
\usepackage{cite}

\newtheoremstyle{style}% name of the style to be used
  {\topsep}% measure of space to leave above the theorem. E.g.: 3pt
  {\topsep}% measure of space to leave below the theorem. E.g.: 3pt
  {\itshape}% name of font to use in the body of the theorem
  {0pt}% measure of space to indent
  {\bfseries}% name of head font
  {:}% punctuation between head and body
  { }% space after theorem head; " " = normal interword space
  {\thmname{#1}\thmnumber{ #2}\thmnote{ (#3)}}

\theoremstyle{style}
\newtheorem{theorem}{Theorem}
\newtheorem{definition}{Definition}
\newtheorem{proposition}{Proposition}
\newtheorem{lemma}{Lemma}

\DeclareMathOperator*{\bigboxplus}{\scalerel*{\boxplus}{()}}

\crefname{equation}{}{}
\crefname{section}{Sec.}{}
\crefname{table}{Tab.}{}
\crefname{algorithm}{Alg.}{}
\crefname{definition}{Def.}{}
\title{\LARGE \bf
Reachability Analysis of ARMAX Models
}
\author{Laura L\"utzow and Matthias Althoff% <-this % stops a space
\thanks{This work has been financially supported by the European Commission project justITSELF under grant number 817629.}% <-this % stops a space
\thanks{L. L\"utzow and M. Althoff are both with the School of Computation, Information and Technology,
        Technical University of Munich, Garching, Germany.
        \{{\tt\small laura.luetzow@tum.de}, {\tt\small althoff@in.tum.de}\}} %
}

\begin{document}
\maketitle
\thispagestyle{empty}
\pagestyle{empty}

%%%%%%%%%%%%%%%%%%%%%%%%%%%%%%%%%%%%%%%%%%%%%%%%%%%%%%%%%%%%%%%%%%%%%%%%%%%%%%%%
\begin{abstract}
Reachability analysis is a powerful tool for computing the set of states or outputs reachable for a system. While previous work has focused on systems described by state-space models, we present the first methods to compute reachable sets of ARMAX models -- one of the most common input-output models originating from data-driven system identification. The first approach we propose can only be used with dependency-preserving set representations such as symbolic zonotopes, while the second one is valid for arbitrary set representations but relies on a reformulation of the ARMAX model. By analyzing the computational complexities, we show that both approaches scale quadratically with respect to the time horizon of the reachability problem when using symbolic zonotopes. To reduce the computational complexity, we propose a third approach that scales linearly with respect to the time horizon when using set representations that are closed under Minkowski addition and linear transformation and that satisfy that the computational complexity of the Minkowski sum is independent of the representation size of the operands. Our numerical experiments demonstrate that the reachable sets of ARMAX models are tighter than the reachable sets of equivalent state space models in case of unknown initial states. Therefore, this methodology has the potential to significantly reduce the conservatism of various verification techniques. 

\end{abstract}

%%%%%%%%%%%%%%%%%%%%%%%%%%%%%%%%%%%%%%%%%%%%%%%%%%%%%%%%%%%%%%%%%%%%%%%%%%%%%%%%
\section{INTRODUCTION}

In recent years, reachability analysis has become popular for verifying autonomous systems subject to uncertainties \cite{althoff2021set}.
By computing the set of states or outputs reachable for a system, we can verify whether unsafe states can be reached. 
This analysis requires conformant system models \cite{roehm2019conformance} to transfer verification results to real systems.

Typically, models must be identified from experiments that provide measured outputs for applied inputs.
Based on this input-output data, the most natural approach to system identification is to assume a model structure where the current output is a function of the previous outputs and the present and previous inputs. These models are called input-output models \cite{ljung2002identification}. Specifically, we will consider ARMAX models in this paper.
As the model is linear in the unknown parameters, the parameters can be identified with well-established regression methods such as linear least squares.
Another popular model structure, which is predominantly used for control applications, is a state-space (SS) model.
In contrast to input-output models, SS models are not solely described by input-output data but contain the system state. 
Since the system state can usually not be measured, the direct identification of SS models from measurements is not straightforward \cite{overschee1996subspace}. 

Although there is a vast amount of literature about data-driven reachability analysis \cite{haesaert2017data,devonport2020data,matavalam2020data,djeumou2021control,chakrabarty2022active,luetzow2023density,alanwar2023data}, most approaches were developed for SS models and require that the initial state or the initial set of states is known. 
To the best knowledge of the authors, there exists no publication that computes the reachable sets of input-output models, although their identification is more straightforward, and they can be initialized solely with measurements.
By investigating reachability methods for ARMAX models, this work closes the gap between data-driven system identification and reachability analysis and therefore enables researchers and engineers to directly use models that are identified from input-output data for safety verification. 

The main contributions of this work are:
\begin{enumerate}
    \item We are the first to develop methods for reachability analysis of ARMAX models. 
    \item For this, we derive a new formulation of the ARMAX model that can be used to predict future outputs without knowledge of intermediate outputs.
    \item We analyze the scalability of our approaches in numerical experiments and compare the computed reachable sets with the reachable sets of equivalent SS models.
\end{enumerate}

The paper is organized as follows: In \cref{sec:prelim}, basic concepts regarding reachability analysis and ARMAX models are introduced. Our novel methods to compute the reachable sets of ARMAX models are presented in \cref{sec:reach}, while their computational complexity is analyzed in \cref{sec:complexity}. In \cref{sec:results}, we demonstrate the accuracy and the scalability of our approaches, and \cref{sec:conclusion} concludes this paper.

%%%%%%%%%%%%%%%%%%%%%%%%%%%%%%%%%%%%%%%%%%%%%%%%%%%%%%%%%%%%%%%%%%%%%%%%%%%%%%%%
\section{PRELIMINARIES} \label{sec:prelim}
\subsection{Notations}
In the remainder of this work, we will denote sets by calligraphic letters, matrices by uppercase letters, and vectors by lowercase letters.
The symbol $\mathbf{0}$ represents a matrix of zeros of appropriate dimensions, and  $\mathbf{I}$ is the identity matrix.
We further introduce the linear transformation of a set $\mathcal{S} \subset \mathbb{R}^{n}$ with a matrix $A \in \mathbb{R}^{m \times n}$, which is defined as $A\mathcal{S} = \{As | s \in \mathcal{S}\}$. 
The Cartesian product of the sets $\mathcal{S}_a\subset \mathbb{R}^{n},~\mathcal{S}_b\subset \mathbb{R}^{m}$ is defined as $\mathcal{S}_a \times \mathcal{S}_b = \{ [s_a^T ~ s_b^T]^T~|~ s_a \in \mathcal{S}_a, ~s_b \in \mathcal{S}_b \}$.
The Minkowski sum of two independent sets $\mathcal{S}_a,~\mathcal{S}_b\subset \mathbb{R}^{n}$ can be computed as $\mathcal{S}_a \oplus \mathcal{S}_b = \{s_a + s_b | s_a \in \mathcal{S}_a,~s_b\in \mathcal{S}_b\}$. 
For the generating function ${S}(\lambda)$ of the set $\mathcal{S}(\Lambda):=\{{S}(\lambda)|\lambda \in \Lambda \}$, the exact addition is defined as
$\mathcal{S}_a(\Lambda) \boxplus \mathcal{S}_b(\Lambda) = \{S_a(\lambda) + S_b(\lambda) | \lambda \in \Lambda\}$ and thus able to consider dependencies between $\mathcal{S}_a$ and $\mathcal{S}_b$.

\subsection{Reachability Analysis}
In reachability analysis, we compute the set of states or outputs that is reachable for a given dynamical system subject to uncertainties. The uncertainties can arise from uncertain initial states $x(0)\in \mathcal{X}(0)$, uncertain inputs $u(k) \in \mathcal{U}(k)$, unknown process disturbances $w(k) \in \mathcal{W}(k)$, and measurement disturbances $v(k) \in \mathcal{V}(k)$ acting on the system. The uncertainty sets can be estimated from measurements by conformant identification algorithms as proposed in \cite{Liu2023conf}.
To define reachable sets, let us first introduce the following model.

\begin{definition} \label{def:ss}\textbf{Linear discrete-time SS model.}
A linear discrete-time SS model is defined by
\begin{subequations} \label{eq:SS}
    \begin{align}
    x(k+1) &= {A} x(k) + {B} u(k) + w(k) \label{eq:SSstate}\\\
    y(k) &= {C} x(k) + {D} u(k) + v(k) \label{eq:SSoutput},    
\end{align}
\end{subequations}
where $x(k) \in \mathbb{R}^{n_x}$ is the system state at time $k$, $y(k) \in \mathbb{R}^{n_{y}}$ is the measured output, $u(k) \in \mathbb{R}^{n_u}$ is the applied input, and $w(k) \in \mathbb{R}^{n_x}$ and $v(k) \in \mathbb{R}^{n_{y}}$ are the process and the measurement disturbance, respectively. The matrices $A$, $B$, $C$ and $D$ have proper dimension.
\end{definition}

\begin{proposition}\textbf{Reachability analysis for SS models.} \label{prop:SS_reach}
The reachable set of linear discrete-time SS models 
can be calculated as 
\begin{align}
    \mathcal{Y}(k) &= {C}{A}^k \mathcal{X}(0) \oplus \bigoplus_{i=1}^k {C}{A}^{i-1} \bigl( {B}\mathcal{U}(k-i) \oplus \mathcal{W}(k-i)\bigr) \notag\\
    &\quad \oplus {D}\mathcal{U}(k) \oplus \mathcal{V}(k). \label{eq:SS_reach}
\end{align}
\begin{proof}
The reachable state set can be computed through a set-based evaluation of \cref{eq:SS}:
\begin{subequations}
\begin{align}
    \mathcal{X}(k+1) &= {A} \mathcal{X}(k) \oplus {B} \mathcal{U}(k) \oplus \mathcal{W}(k) \label{eq:SSstate_set} \\ 
    \mathcal{Y}(k) &= {C}\mathcal{X}(k) \oplus {D}\mathcal{U}(k) \oplus  \mathcal{V}(k). \label{eq:SSoutput_set}    
\end{align}
\end{subequations}
By recursively applying \cref{eq:SSstate_set} and inserting the result in \cref{eq:SSoutput_set}, we obtain the formula in \cref{eq:SS_reach}.
\end{proof}
\end{proposition}

The complexity and accuracy of these computations depend on the chosen set representation. In this work, we will use symbolic zonotopes to demonstrate the proposed reachability methods.

\subsection{Zonotopes}
A zonotope is a convex set representation whose representation size scales well with the system dimension \cite{girard2005zonotopes}. 

\begin{definition} \textbf{Zonotopes \cite{kuehn1998wrapping}.} \label{def:zon}
    A zonotope $\mathcal{Z}$ of order $o$ can be described by
\begin{align*}
    \mathcal{Z} = \left\{c+\sum_{i=1}^{on} \lambda_i g_i \middle|\lambda_i\in[-1,1]\right\}=\langle  c,G \rangle,
\end{align*}
where $c \in \mathbb{R}^{n}$ is the zonotope center and $G=[g_1~...~g_{on}] \in \mathbb{R}^{n \times {on}}$ is the generator matrix.
\end{definition}

Since zonotopes are not able to consider dependencies when adding set-valued variables, we adopt the idea of symbolic zonotopes: 

\begin{definition} \textbf{Symbolic zonotopes \cite{combastel2020symb}.}
A symbolic zonotope $\mathcal{Z}^*$ of order $o$ is defined as
\begin{align*}
    \mathcal{Z}^* 
    = \left\{c+\sum_{i=1}^{on} \lambda_{l(i)} g_{l(i)} \middle|\lambda_{l(i)}\in[-1,1]\right\}
    = \left\langle c,G^{|l}\right\rangle^*
\end{align*}
where $c \in \mathbb{R}^{n}$ is the zonotope center and $\lambda_{l(i)}$ is the symbolic zonotope factor that is uniquely identified by the positive integer $l(i)$.
$G^{|l}=[g_{l(1)}~...~g_{l({on})}] \in \mathbb{R}^{n \times {on}}$ is the generator matrix whose columns $g_{l(\cdot)}$ are labeled by the vector $l=[l(1)~...~l({on})]\in \mathbb{N}^{on}$.
\end{definition}

By ensuring that dependent generators from different sets have identical labels $l(i)$ and thus identical zonotope factors $\lambda_{l(i)}$, dependencies between sets can be preserved.
New labels are only generated when creating independent generators. 
Obviously, a symbolic zonotope $\mathcal{Z}^*=\langle c,G^{|l}\rangle^*$ whose generators are independent of any other set can be equivalently described by the zonotope $\mathcal{Z}=\langle c,G\rangle$. 

Before introducing mathematical operations on symbolic zonotopes, we need to define the addition operation for labeled matrices:

\begin{definition}\textbf{Addition of labeled matrices\cite[Def. 4.10]{combastel2020symb}.}
The sum of two labeled matrices, $M_a^{|l_a}$ and $M_a^{|l_a}$, is defined as
\begin{align*}
    M_a^{|l_a}+M_b^{|l_b} = [M_a^{|l_a\backslash l_b}~~(M_a^{|l_a\cap l_b}+M_b^{|l_b\cap l_a})~~M_b^{|l_b\backslash l_a}],
\end{align*}
i.e., columns with identical labels are summed together; other columns are concatenated while their labels are transferred to the new matrix. 
\end{definition}

We can use the following mathematical operations on symbolic zonotopes \cite[Sec. 6]{combastel2020symb}:
Given two symbolic zonotopes $\mathcal{Z}_a^*= \langle c_a,G_a^{|l_a}\rangle^*$ and $\mathcal{Z}_b^*= \langle c_b,G_b^{|l_b}\rangle^*$,
the exact addition can be computed as $\mathcal{Z}_a^* \boxplus  \mathcal{Z}_b^*=\langle c_a+c_b,G_a^{|l_a}+G_b^{|l_b}\rangle^*$.
Exact addition is equal to the Minkowski sum if $\mathcal{Z}_a^*$ and $\mathcal{Z}_b^*$ are independent, i.e., the labels $l_a$ and $l_b$ are mutually different, which leads to $\mathcal{Z}_a^* \oplus  \mathcal{Z}_b^*=\langle c_a+c_b,[G_a~G_b]^{|[l_a~ l_b]}\rangle^*$. 
The linear transformation of $\mathcal{Z}_a^*$ with a matrix $A$ can be computed as $A\mathcal{Z}_a^*=\langle Ac_a,(AG_a)^{|l_a}\rangle^*$. 
The Cartesian product of $\mathcal{Z}_a^*$ and $\mathcal{Z}_b^*$ is
\begin{align*}
    \mathcal{Z}_a^* \times \mathcal{Z}_b^* = \left\langle\begin{bmatrix}
        c_a \\ c_b
    \end{bmatrix},\begin{bmatrix}
        G_a \\ \mathbf{0}
    \end{bmatrix}^{|l_a} + \begin{bmatrix}
        \mathbf{0} \\ G_b
    \end{bmatrix}^{|l_b}\right\rangle^*.
\end{align*}

\subsection{ARMAX Models}

As discussed in the introduction, input-output models are preferred over SS models in applications that require a preceding or simultaneous system identification.  
One popular representative of input-output models is the ARMAX model:

\begin{definition} \textbf{ARMAX model \cite{isermann2010identification}.}
The Autoregressive Moving Average model with exogenous Input (ARMAX model) can be described by   
\begin{align}
    y(k) &= \sum_{i=1}^p \bar{A}_{i} y(k-i) + \sum_{i=0}^p \bar{B}_{i} \tilde{u}(k-i) \label{eq:ARMAXnormal}
\end{align}
where 
\begin{align*}
    \tilde{u}(k-i) = \begin{bmatrix}
        u(k-i) \\ w(k-i) \\ v(k-i)
    \end{bmatrix},
\end{align*}
$\bar{A}_{i}\in \mathbb{R}^{n_{y} \times n_{y}}$,
 $\bar{B}_{i}\in \mathbb{R}^{n_{y} \times n_{\tilde{u}}}$, and $n_{\tilde{u}}=n_u+n_w+n_v$.
\end{definition}

In contrast to standard notations, which contain an input and a disturbance term (e.g., see \cite[p.83]{ljung2002identification}), we consider the combined input and disturbance $\tilde{u}$. This results from the conversion from SS models, which is explained subsequently.

\begin{proposition}  \textbf{From SS to ARMAX models.}\label{prop:transfSS_ARMAX}
The discrete-time SS model from \cref{eq:SS}
can be converted to an ARMAX model of the form in \cref{eq:ARMAXnormal} with
\begin{subequations} \label{eq:ARMAX_param}
    \begin{align}
    & &&\bar{A}_{i} = -{C}({A}+M{C})^{i-1}M, \label{eq:ARMAX_paramA}\\
    \bar{B}_{0} &= D_{\tilde{u}}, &&\bar{B}_{i} = {C}({A}+M{C})^{i-1}({B_{\tilde{u}}}+M{D_{\tilde{u}}}), \label{eq:ARMAX_paramB}
\end{align}
\end{subequations}
where $i=1,...,p$,  $D_{\tilde{u}}=[{D}~\mathbf{0}~\mathbf{I}]$, $B_{\tilde{u}}=[{B}~\mathbf{I}~\mathbf{0}]$, and $M\in \mathbb{R}^{n_x \times n_{y}}$ is the deadbeat observer gain that satisfies $({A}+M{C})^k= \mathbf{0}$ for $k\geq p$. 

\begin{proof}
See \cite{phan1993linear} for the SS model
\begin{align*}
    x(k+1) &= {A} x(k) + B_{\tilde{u}} \tilde{u}(k)\\
    y(k) &= {C} x(k) + D_{\tilde{u}} \tilde{u}(k),    
\end{align*}
which is obtained from \cref{eq:SS} by combining the inputs and disturbances in $\tilde{u}$.
\end{proof}
\end{proposition}
For high-dimensional systems where the computational complexity of estimating the deadbeat observer gain $M$ is too big, an equivalent ARMAX model can be computed using the Cayley-Hamilton theorem \cite[Sec. 2.2.2]{aoki1991arma2ss}. This transformation method scales well to high dimensions and leads to diagonal $\bar{A}_i$ matrices but generally results in an increased model order $p$.
For the inverse problem of finding an SS formulation of a given ARMAX model, the partial fraction expansion \cite[Sec. 2.1.2]{aoki1991arma2ss} can be used.

\subsection{Problem Statement}
In this work, we compute the set of possible measurements $\mathcal{Y}(k)$, with $k \geq p$, for a system described by an ARMAX model. We assume that the initial measurements $y(i)$, with $i=0,...,p-1$, such as the input sets $\mathcal{U}(i)$, and the set of disturbances $\mathcal{W}(i)$ and $\mathcal{V}(i)$, with $i=0,...,k$, are given.

%%%%%%%%%%%%%%%%%%%%%%%%%%%%%%%%%%%%%%%%%%%%%%%%%%%%%%%%%%%%%%%%%%%%%%%%%%%%%%%%

\section{REACHABILITY ANALYSIS FOR ARMAX MODELS} \label{sec:reach}
In this section, we present our approaches to compute the reachable set of ARMAX models. First, we will calculate the reachable set using dependency-preserving set representations, followed by the derivation of a reachability algorithm that is valid for arbitrary set representations. Lastly, we will propose an approach that improves the scalability w.r.t. the time horizon of the reachability problem.

\subsection{Reachability Analysis with Dependency-preserving Sets} \label{sec:meth1}

For dependency-preserving set operations, such as symbolic zonotopes, we can directly evaluate \cref{eq:ARMAXnormal} in a set-based fashion:

\begin{proposition}\label{theo:reach_zon}\textbf{Reachability analysis with dependency-preserving sets.}
By representing the combined input and disturbance sets $\tilde{\mathcal{U}}(i)={\mathcal{U}}(i) \times {\mathcal{W}}(i) \times {\mathcal{V}}(i)$, $0\leq i\leq k$, with a dependency-preserving set representation, 
we can compute the reachable set of \cref{eq:ARMAXnormal} as
\begin{align}
    \mathcal{Y}(k) &= \bigboxplus_{i=1}^p \bar{A}_{i} \mathcal{Y}(k-i) \boxplus \bigboxplus_{i=0}^p \bar{B}_{i} \tilde{\mathcal{U}}(k-i), \label{eq:ARMAXzon_reach}
\end{align}
where the output sets $\mathcal{Y}(i)$, $i=0,...,p-1$, can be initialized from the measurements $y(i)$.

\begin{proof}
By keeping track of the dependencies between the sets, the result of \cref{eq:ARMAXzon_reach} is exact.
\end{proof}
\end{proposition}

Using Minkowski addition instead of exact addition would neglect the dependencies between the summed sets, allowing the same variable to have different values in different occurrences.
This dependency problem \cite{kochdumper2020dependecies,mitchell2019invariant,gruber2021mpc,jaulin2001interval} would lead to overapproximative set computations as later visualized in \cref{sec:results}.

A reachability algorithm that can be used with set representations that are not dependency-preserving is described in the following section.

\subsection{Reachability Analysis with Arbitrary Set Representations} \label{sec:meth2}

As many set representations cannot consider dependencies between summed variables, we reformulate \cref{eq:ARMAXnormal} as a sum over independent variables.

\begin{theorem} \label{prop:ARMAXreform}\textbf{Alternative ARMAX formulation.} 
The ARMAX model from \cref{eq:ARMAXnormal} can be formulated as 
\begin{align}
    \tilde{y}_{(k:k_+)} &= {\tilde{A}(k) \tilde{y}_\text{init}} +  \sum_{i=0}^{k_+}\tilde{B}_{i}(k) \tilde{u}(k_+-i) \label{eq:ARMAXtvp}
\end{align}
with the stacked outputs
\begin{align}
    \tilde{y}_\text{init}= \begin{bmatrix}
        y(0) \\
        y(1) \\
        \vdots\\
        y(p-1)
    \end{bmatrix},~ 
    \tilde{y}_{(k:k_+)}= \begin{bmatrix}
        y(k) \\
        y(k+1) \\
        \vdots\\
        y(k_+)
    \end{bmatrix}\in \mathbb{R}^{pn_{y}}, \label{eq:ytilde}    
\end{align}
and $k_+ = k+p-1$. 
The time-varying parameters $\tilde{A}(k) \in \mathbb{R}^{pn_{y} \times pn_{y}}$ 
and $\tilde{B}_{i}(k) \in \mathbb{R}^{pn_{y} \times n_{\tilde{u}}}$ 
can be computed as
\begin{subequations}\label{eq:ARMAXtvp_param}
    \begin{align}    
    \tilde{A}(k) &= \bar{A}_\text{ext}^{k}, \label{eq:ARMAXtvp_paramA} \\
    \tilde{B}_{i}(k) &=  \sum_{j=0}^{k-1} \bar{A}_\text{ext}^{j} \bar{B}_{\text{ext},i-j},  
    \label{eq:ARMAXtvp_paramB}
\end{align}
\end{subequations}
where 
\begin{align}  
    \bar{A}_\text{ext} &= \begin{bmatrix}  
        \mathbf{0} &\mathbf{I} & \mathbf{0} & ... & \mathbf{0}  \\
        \mathbf{0} &\mathbf{0} & \mathbf{I} & ... & \mathbf{0}  \\
        \vdots & \vdots & \vdots  & \ddots & \vdots \\
        \mathbf{0} & \mathbf{0} & \mathbf{0} & ... & \mathbf{I} \\
        \bar{A}_{p} & \bar{A}_{p-1} &  \bar{A}_{p-2} & ...&  \bar{A}_{1}
    \end{bmatrix},%\label{eq:Atilde}
    ~ \bar{B}_{\text{ext},i} = \begin{bmatrix}
        \mathbf{0} \\
        \mathbf{0} \\
        \vdots\\
        \mathbf{0}       \\
        \bar{B}_{i} 
    \end{bmatrix}\label{eq:Btilde}  ,
\end{align}
and $\bar{B}_{i}=\mathbf{0}$ for $i< 0$ or $i>p$. 

\begin{proof}
By defining the stacked output and the extended matrices $\bar{A}_\text{ext}$ and $\bar{B}_{\text{ext},i}$ as in \cref{eq:ytilde,eq:Btilde},
we can rewrite the ARMAX model from \cref{eq:ARMAXnormal} as 
\begin{align}
    \tilde{y}_{(k:k_+)} &= \bar{A}_\text{ext} \tilde{y}_{(k-1,k_+-1)} +  \sum_{i=0}^p  \bar{B}_{\text{ext},i} \tilde{u}(k_+-i).  \label{eq:ARMAXext}
\end{align}

After recursively evaluating \cref{eq:ARMAXext}, we obtain a formulation where $\tilde{y}_{(k:k_+)}$ only depends on the initial measurements {$\tilde{y}_\text{init}:=\tilde{y}_{(0,p-1)}$} and all past inputs:
\begin{align*}
    \tilde{y}_{(k:k_+)} &= {\bar{A}_\text{ext}^{k}} \tilde{y}_\text{init} +  {\sum_{j=0}^{k-1} \bar{A}_\text{ext}^{j} \sum_{i=0}^p \bar{B}_{\text{ext},i} \tilde{u}(k_+-j-i)}.
\end{align*}
The parameters $\tilde{B}_{i}(k)$ can be obtained by rearranging the second summand such that each variable $\tilde{u}$ only occurs once.
We start with shifting the summation index $i$ by $+j$:
\begin{align*}
    \tilde{y}_{(k:k_+)} &= {\bar{A}_\text{ext}^{k}} \tilde{y}_\text{init} + {\sum_{j=0}^{k-1} \bar{A}_\text{ext}^{j} \sum_{i=j}^{p+j} \bar{B}_{\text{ext},i-j} \tilde{u}(k_+-i)}.
\end{align*}
By defining $\bar{B}_{\text{ext},i-j}$ to be the zero matrix for $i<j$ and $i>p+j$, we can insert the minimum and maximum value of $j$ in the summation limits of the second sum: 
\begin{align*}
    \tilde{y}_{(k:k_+)} &= {\bar{A}_\text{ext}^{k}} \tilde{y}_\text{init} + {\sum_{j=0}^{k-1} \bar{A}_\text{ext}^{j} \sum_{i=0}^{k_+} \bar{B}_{\text{ext},i-j} \tilde{u}(k_+-i)}.
\end{align*}
Switching the summation order to
\begin{align*}    
    \tilde{y}_{(k:k_+)} &= {\bar{A}_\text{ext}^{k}} \tilde{y}_\text{init} +\sum_{i=0}^{k_+} {\sum_{j=0}^{k-1} \bar{A}_\text{ext}^{j} \bar{B}_{\text{ext},i-j}} \tilde{u}(k_+-i)
\end{align*}
results in \cref{eq:ARMAXtvp} and \cref{eq:ARMAXtvp_param}.
By a case distinction in $i$, we could exclude the summands in the sum over $j$ where $\bar{B}_{\text{ext},i-j}=\mathbf{0}$, but this is not done here due to notational simplicity. 
\end{proof} 
\end{theorem}

As we do not have any dependencies between the variables in the reformulated ARMAX model of \cref{prop:ARMAXreform}, we can directly compute the reachable sets by a set-based evaluation of \cref{eq:ARMAXtvp}: 
\begin{align}
    \tilde{\mathcal{Y}}_{(k:k_+)} &=   \tilde{A}(k) \tilde{y}_\text{init} \oplus \bigoplus_{i=0}^{k_+} \tilde{B}_{i}(k) \tilde{\mathcal{U}}(k_+-i) \label{eq:ARMAXtvp_reach1}
\end{align}
where 
\begin{align}\label{eq:cartProdY}
    \tilde{\mathcal{Y}}_{(k:k_+)} = \mathcal{Y}(k) \times \mathcal{Y}(k+1) \times \cdots \times \mathcal{Y}(k_+).
\end{align}
However, each evaluation requires $(k+1)$ Minkowski sums and linear transformations of sets, which can be expensive when computing the reachable sets for a large number of time steps.
Thus, we propose a more general approach that can reuse previous results. We need the following lemmas:

\begin{lemma}\label{lem:recParam}
The time-varying parameters $\tilde{A}(k+\Delta k)$ and $\tilde{B}_{i}(k+\Delta k)$, $\Delta k\geq 0$, can be
computed from $\tilde{A}(k)$ and $\tilde{B}_{i}(k)$ for $i \geq p + \Delta k$ with   
\begin{subequations}\label{eq:ARMAXtvp_paramRec}
    \begin{align}
    \tilde{A}(k+\Delta k) &= \bar{A}_{\text{ext}}^{\Delta k}  \tilde{A}(k), \label{eq:ARMAXtvp_paramRecA}\\
    \tilde{B}_{i}(k+\Delta k) &= \bar{A}_{\text{ext}}^{\Delta k} \tilde{B}_{i-\Delta k}(k).
    \label{eq:ARMAXtvp_paramRecB}
\end{align}
\end{subequations}

\begin{proof}
Equation \cref{eq:ARMAXtvp_paramRecA} follows directly from \cref{eq:ARMAXtvp_paramA}.
Equation \cref{eq:ARMAXtvp_paramRecB} can be obtained from \cref{eq:ARMAXtvp_paramB} by computing $\tilde{B}_{i}(k+\Delta k)$ for the case $i\geq p+\Delta k$. Since $\bar{B}_{\text{ext},i-j}$ is the zero matrix for $i-j>p$, we only have to consider summands with $j\geq \Delta k$, which leads to
    \begin{align*}    
        \tilde{B}_{i}(k+\Delta k) &= \sum_{j=\Delta k}^{k+\Delta k-1} \bar{A}_\text{ext}^{j} \bar{B}_{\text{ext},i-j}\\
        &= \sum_{j=0}^{k-1} \bar{A}_\text{ext}^{j+\Delta k} \bar{B}_{\text{ext},i-\Delta k-j} = \bar{A}_\text{ext}^{\Delta k}\tilde{B}_{i-\Delta k}(k).
    \end{align*}\vspace{-0.5cm}
    
\end{proof}
\end{lemma}

\begin{lemma}\label{lem:recParam2} For $i< k$ and $\Delta k \geq 0$, we can directly reuse old parameters with
\begin{align}
    \tilde{B}_{i}(k+\Delta k) &= \tilde{B}_{i}(k). \label{eq:ARMAXtvp_paramRecB2}
\end{align}
\begin{proof}
We evaluate \cref{eq:ARMAXtvp_paramB} for the case $i < k$. Since $\bar{B}_{\text{ext},i-j}=\mathbf{0}$ for $i-j<0$, we can remove summands with $j>i$, which results in
    \begin{align*}    
        {\tilde{B}_{i}(k)} &= \sum_{j=0}^{i} \bar{A}_\text{ext}^{j} \bar{B}_{\text{ext},i-j}.
    \end{align*}
    As this expression is independent of $k$, the parameters $\tilde{B}_{i}$ will not change at future time steps $k+\Delta k$.
\end{proof}
\end{lemma}

\begin{lemma}\label{lem:recParam3} The parameters $\tilde{B}_{i+1}(k)$ can be obtained from $\tilde{B}_{i}(k)$ with
\begin{align}
    \tilde{B}_{i+1}(k) &= \bar{A}_\text{ext}\tilde{B}_{i}(k) + \bar{B}_{\text{ext},i+1} - \bar{A}_\text{ext}^{k}\bar{B}_{\text{ext},i+1-k}. \label{eq:ARMAXtvp_paramRecB3}
\end{align}
\begin{proof}
We compute $\tilde{B}_{i+1}(k)$ with \cref{eq:ARMAXtvp_paramB} and shift the summation index $j$ by $-1$:
    \begin{align*}    
        {\tilde{B}_{i+1}(k)} &= \sum_{j=-1}^{k-2} \bar{A}_\text{ext}^{j+1} \bar{B}_{\text{ext},i-j}\\
        &=\bar{A}_\text{ext} \bigl(\sum_{j=0}^{k-1} \bar{A}_\text{ext}^{j} \bar{B}_{\text{ext},i-j} + \sum_{j=-1}^{-1} \bar{A}_\text{ext}^{j} \bar{B}_{\text{ext},i-j} \\ &\quad- \sum_{j=k-1}^{k-1} \bar{A}_\text{ext}^{j} \bar{B}_{\text{ext},i-j}\bigr).
    \end{align*}
    Replacing the first sum with $\tilde{B}_{i}(k)$ results in \cref{eq:ARMAXtvp_paramRecB3}.
\end{proof}
\end{lemma}

\begin{theorem}\label{theo:reach_gen}\textbf{Reachability analysis with arbitrary sets.}
The reachable sets of an ARMAX model at the consecutive time points $k=k_{\text{init}},~k_{\text{init}}+1,~ ....,~p+k_h$, where $k_{\text{init}} \geq p$, can be obtained with \cref{alg:reachRec}.

\begin{proof}
    Inserting \cref{algline:ARMAXtvp_S} in \cref{algline:ARMAXtvp_reach} shows the equality to \cref{eq:ARMAXtvp_reach1}.
    By computing the Cartesian product $\tilde{\mathcal{Y}}_{(k:k_+)}$ as defined in \cref{eq:cartProdY}, we obtain $p$ reachable sets in each iteration, and thus, the time step $k$ can be incremented by $p$ after evaluating \cref{algline:ARMAXtvp_reach}.
    The parameters $\tilde{B}_i(k)$, $i=0,...,p-1$, which are used in \cref{algline:ARMAXtvp_reach}, do not change after their initialization at $k=k_{\text{init}}\geq p$ since all indices $i$ satisfy the condition $i< k$ of \cref{lem:recParam2}.
    In contrast, the parameters $\tilde{B}_i(k)$, $i=p,...,2p-1$, which are required for \cref{algline:ARMAXtvp_SRec}, do not satisfy this conditions if the updated $k$ is smaller than $3p$. They are recomputed in \cref{algline:ARMAXtvp_paramB2p,algline:ARMAXtvp_paramB2i}. 
\begin{table}[h]
    \vspace{-0.15cm}
\end{table}
\begin{algorithm}[h]
\caption{Reachability analysis with arbitrary sets.}\label{alg:reachRec}
\begin{algorithmic}[1]
\State $k \gets k_{\text{init}},\quad k_+ \gets k_{\text{init}} +p-1$ 
\State $\tilde{A} \gets$ \cref{eq:ARMAXtvp_paramA} \label{algline:Atilde}
\State $\tilde{B}_0 \gets$ \cref{eq:ARMAXtvp_paramB}\label{algline:Btilde0}
\State $\tilde{B}_i,$~{\small $i=1,...,k_+$} $\gets$ \cref{eq:ARMAXtvp_paramRecB3}\label{algline:Btildei}
\State ${\mathcal{S}} \gets \tilde{A} \tilde{y}_\text{init} \oplus \bigoplus_{i=p}^{k_+} \tilde{B}_{i} \tilde{\mathcal{U}}(k_+-i)$ \label{algline:ARMAXtvp_S}
\While{$k\leq p+k_h$}
\State $\tilde{\mathcal{Y}}_{(k:k_+)} \gets{\mathcal{S}}\oplus \bigoplus_{i=0}^{p-1}  \tilde{B}_{i} \tilde{\mathcal{U}}(k_+-i)$ \label{algline:ARMAXtvp_reach} 
\State $k \gets k+p,\quad k_+ \gets k_+ +p$
\If{$k< 3p$} \label{algline:test}
\State $\tilde{B}_p \gets$ \cref{eq:ARMAXtvp_paramB}\label{algline:ARMAXtvp_paramB2p} 
\State $\tilde{B}_i$, {\small $i=p+1,...,2p-1$} $\gets$ \cref{eq:ARMAXtvp_paramRecB3} \label{algline:ARMAXtvp_paramB2i}
\EndIf
\State ${\mathcal{S}} \gets \bar{A}_\text{ext}^{p} {\mathcal{S}}  \oplus \bigoplus_{i=p}^{2p-1} \tilde{B}_{i} \tilde{\mathcal{U}}(k_+-i)$  \label{algline:ARMAXtvp_SRec}
\EndWhile
\end{algorithmic}
\end{algorithm}
    
    To obtain the recursive formula in \cref{algline:ARMAXtvp_SRec}, we compare ${\mathcal{S}}$, computed with  \cref{algline:ARMAXtvp_S}, for the time steps $k-p$ and $k$:
    \begin{align*}
        {\mathcal{S}}(k-p) &=   \tilde{A}(k-p) \tilde{y}_\text{init} \oplus \bigoplus_{i=p}^{k-p} \tilde{B}_{i}(k-p) \tilde{\mathcal{U}}(k-p-i).
    \end{align*}
    With a case distinction in $i$ and \cref{lem:recParam} using $\Delta k = p$, ${\mathcal{S}}(k)$ can be written as
    \begin{align*}
        {\mathcal{S}}(k) &= {\bar{A}_\text{ext}^{p}  \tilde{A}(k-p) \tilde{y}_\text{init} \oplus \bigoplus_{i=2p}^{k} \bar{A}_\text{ext}^{p} \tilde{B}_{i-p}(k-p) {\tilde{\mathcal{U}}}(k-i)} \\ 
        & \qquad \oplus \bigoplus_{i=p}^{2p-1} \tilde{B}_{i}(k) \tilde{\mathcal{U}}(k-i).
    \end{align*}
   With a shift of the first summation index $i$ by $-p$, the first line equals $\bar{A}_\text{ext}^{p}{\mathcal{S}}(k-p)$, which leads to \cref{algline:ARMAXtvp_SRec}.
\end{proof}
\end{theorem}

In each iteration, \cref{alg:reachRec} requires $(2p+1)$ Minkowski sums and linear set transformations to obtain $\tilde{\mathcal{Y}}_{(k:k_+)}$ and is thus computationally more efficient than \cref{eq:ARMAXtvp_reach1} if we calculate multiple consecutive reachable sets.

\subsection{Reachability Analysis with Reduced Complexity}
When using symbolic zonotopes as set representation, the number of generators of ${\mathcal{Y}}$ in \cref{theo:reach_zon} and of ${\mathcal{S}}$ in \cref{theo:reach_gen} increases with the time step $k$. Due to the linear transformations that are applied to ${\mathcal{Y}}$ and ${\mathcal{S}}$ at each iteration, \cref{theo:reach_zon,theo:reach_gen} scale quadratically with the time horizon $k_h$  if we compute the reachable sets for $k=p,...,p+k_h$ (see \cref{sec:complexity} for a detailed analysis). 
By upper-bounding the number of generators of ${\mathcal{Y}}$ and ${\mathcal{S}}$ using order reduction methods, the complexity of \cref{theo:reach_zon,theo:reach_gen} can be reduced to $\mathcal{O}(k_h)$ with the downside that the reachable sets will not be exact anymore and prone to the wrapping effect.
In comparison, classical reachability analysis for SS models with zonotopes scales also quadratically w.r.t. $k_h$, but the complexity can be reduced to $\mathcal{O}(k_h)$ without wrapping effect by rescheduling the computations as proposed in \cite{girard2006efficient}.
This rescheduling approach can be adapted to ARMAX models in the following way:

\begin{theorem} \textbf{Efficient reachability analysis.} \label{theo:reach_eff}
    If the combined input and disturbance sets can be described by the constant set $\tilde{\mathcal{U}}_c$ and the time-varying signal $\tilde{u}_v(i)$ such that $\tilde{\mathcal{U}}(i) = \tilde{\mathcal{U}}_c + \tilde{u}_v(i)$, $0 \leq i\leq p+k_h$, we can compute the reachable sets at the consecutive time points $k_{\text{init}},~k_{\text{init}}+1,~ ....,~p+k_h$, where $k_{\text{init}} \geq p$, with \cref{alg:approx}.
\vspace{-0.2cm}
\begin{algorithm}[hbt]
\caption{Efficient reachability analysis.}\label{alg:approx}
\begin{algorithmic}[1]
\State $k \gets k_{\text{init}},\quad k_+ \gets k_{\text{init}} +p-1$ 
\State $\tilde{A} \gets$ \cref{eq:ARMAXtvp_paramA} \label{algline:A}
\State $\tilde{B}_0 \gets$ \cref{eq:ARMAXtvp_paramB}\label{algline:B}
\State $\tilde{B}_i,$~{\small $i=1,...,k_+$} $\gets$ \cref{eq:ARMAXtvp_paramRecB3}\label{algline:Bi}
\State $\mathcal{S}_c \gets \bigoplus_{i=0}^{k-1}\tilde{B}_i \tilde{\mathcal{U}}_c$ \label{algline:Sc}
\State $\mathcal{T}_{c1} \gets \bigoplus_{i=k}^{k_+}\tilde{B}_i \tilde{\mathcal{U}}_c$  \label{algline:Tc1}
\State $s_v \gets \tilde{A} \tilde{y}_\text{init} + \sum_{i=p}^{k_+} \tilde{B}_{i}\tilde{u}_v(k_+-i)$\label{algline:sv}
\While{$k\leq p+k_h$}
\State $\tilde{\mathcal{Y}}_c \gets \mathcal{S}_c \oplus \mathcal{T}_{c1}$\label{algline:Yc}
\State $\tilde{y}_v \gets s_v + \sum_{i=0}^{p-1}  \tilde{B}_{i} \tilde{u}_v(k_+-i)$\label{algline:yv}
\State $\tilde{\mathcal{Y}}_{(k:k_+)} \gets \tilde{y}_v \oplus \tilde{\mathcal{Y}}_c$ \label{algline:Ysplit}
\State $k \gets k+p$,\quad $k_+ \gets k_++p$
\If{$k< 3p$} \label{algline:test2} 
\State $\tilde{B}_p \gets$ \cref{eq:ARMAXtvp_paramB} \label{algline:Btilde2p}~ \State $\tilde{B}_i$, {\small $i=p+1,...,2p-1$} $\gets$ \cref{eq:ARMAXtvp_paramRecB3}\label{algline:Btilde2i}
\EndIf
\If{$k=k_{\text{init}}+p$}
\State $\mathcal{T}_{c2} \gets \bigoplus_{i=k-p}^{k_+-p}\tilde{B}_i \tilde{\mathcal{U}}_c$\label{algline:Tc2}
\EndIf
\State $\mathcal{S}_c \gets \mathcal{S}_c \oplus \mathcal{T}_{c2}$ \label{algline:Sc_rec}
\State $\mathcal{T}_{c1} \gets \bar{A}_\text{ext}^p \mathcal{T}_{c1}$ \label{algline:Tc1_rec}
\State $\mathcal{T}_{c2} \gets \bar{A}_\text{ext}^p \mathcal{T}_{c2}$ \label{algline:Tc2_rec}
\State $s_v \gets \bar{A}_\text{ext}^{p} s_v  + \sum_{i=p}^{2p-1} \tilde{B}_{i} \tilde{u}_v(k_+-i)$ \label{algline:sv_rec}
\EndWhile
\end{algorithmic}
\end{algorithm}
    \vspace{-0.4cm}
\begin{proof}
    By using \cref{eq:ARMAXtvp_reach1} with the input sets $\tilde{\mathcal{U}}(i) = \tilde{\mathcal{U}}_c + \tilde{u}_v(i)$, we can split $\tilde{\mathcal{Y}}_{(k:k_+)}$ in \cref{algline:Ysplit} in
    \begin{align*}
    \tilde{\mathcal{Y}}_{(k:k_+)} &=   \underbrace{\left(\tilde{A}(k) \tilde{y}_\text{init} + \sum_{i=0}^{k_+} \tilde{B}_{i}(k) \tilde{u}_v(k_+-i)\right)}_{\tilde{y}_{v(k:k_+)}} \\ & \qquad \oplus \underbrace{\bigoplus_{i=0}^{k_+} \tilde{B}_{i}(k) \tilde{\mathcal{U}}_c}_{\tilde{\mathcal{Y}}_{c(k:k_+)}}.
    \end{align*}
    The vector $\tilde{y}_{v(k:k_+)}$ can be computed analogously to $\tilde{\mathcal{Y}}_{(k:k_+)}$ in \cref{alg:reachRec} using the auxiliary variable $s_{v(k:k_+)}$ (see \cref{algline:sv,algline:sv_rec,algline:yv}).
    To efficiently compute the set $\tilde{\mathcal{Y}}_{c(k:k_+)}$ in \cref{algline:Yc}, we split the Minkowski sum in
    \begin{align*}
    \tilde{\mathcal{Y}}_{c(k:k_+)} &=  \underbrace{\bigoplus_{i=0}^{k-1} \tilde{B}_{i}(k) \tilde{\mathcal{U}}_c}_{\mathcal{S}_c(k)} \oplus \underbrace{\bigoplus_{i=k}^{k_+} \tilde{B}_{i}(k) \tilde{\mathcal{U}}_c}_{\mathcal{T}_{c1}(k)},
    \end{align*}
    where $\mathcal{S}_c(k)$ and $\mathcal{T}_{c1}(k)$ are initialized in \cref{algline:Sc,algline:Tc1}.
    In \cref{algline:Sc_rec,algline:Tc1_rec}, we use \cref{lem:recParam2} to update $\mathcal{S}_c(k)$ and \cref{lem:recParam} to update $\mathcal{T}_{c1}(k)$ as
    \begin{align*}
        \mathcal{S}_c(k+p) &= \mathcal{S}_c(k) \oplus \underbrace{\bigoplus_{i=k}^{k_+} \tilde{B}_{i}(k+p) \tilde{\mathcal{U}}_c}_{\mathcal{T}_{c2}(k)},\\
        \mathcal{T}_{c1}(k+p) &= \bar{A}_\text{ext}^p \mathcal{T}_{c1}(k),
    \end{align*}
    where the auxiliary set $\mathcal{T}_{c2}(k)$ is initialized in \cref{algline:Tc2} with the parameters $\tilde{B}_i(k_{\text{init}}+p)$ and updated recursively with  $\mathcal{T}_{c2}(k+p) = \bar{A}_\text{ext}^p \mathcal{T}_{c2}(k)$ in  \cref{algline:Tc2_rec}. 
\end{proof}
\end{theorem}
As shown in the next section, the computational complexity of this algorithm scales linearly with the time horizon $k_h$ if $\tilde{\mathcal{U}_c}(\cdot)$ is described by a set representation that is closed under Minkowski sum and linear transformation and that satisfies that the computational complexity of the Minkowski sum is independent of the representation size of the operators \cite{girard2006efficient}. These properties are, for example, satisfied by zonotopes or symbolic zonotopes.

\section{Computational Complexity}
\label{sec:complexity}
In this section, we analyze the complexity of computing the reachable sets at the time points $k=p, ..., p+k_h$ with \cref{theo:reach_zon,theo:reach_gen,theo:reach_eff} while representing $\tilde{\mathcal{U}}(\cdot)$ with symbolic zonotopes of order $o$ with ${on_{\tilde{u}}}$ generators (for \cref{theo:reach_gen,theo:reach_eff} we could equivalently use zonotopes). The input and output dimension are described by the nominal input and output dimension $n_{y0}$ and $n_{\tilde{u}0}$ and the scaling factor $f_n$, i.e., $n_{y}=f_n n_{y0}$ and $n_{\tilde{u}} = f_n n_{\tilde{u}0}$. Since \cref{theo:reach_gen,theo:reach_eff} compute $p$ reachable sets at each iteration, we consider the time horizon $k_h$ as a function of $p$, i.e., $k_h = f_k p$.

The dominating complexities w.r.t. the time horizon parameter $f_k$, the scaling factor for the system dimension $f_n$, and the model order $p$ are computed in the remaining part of this section and summarized in \cref{tab:complexity}. A more elaborate version of the computations can be found in the complementary documentation folder in the CORA toolbox \cite{althoff2015introduction}. Since the computational complexity of unary operations is usually neglectable, we consider only binary operations. The complexities of operations on zonotopes are taken from \cite[Tab. 1]{althoff2021set}. Furthermore, we assume the use of textbook methods, not considering special numerical tricks.

\subsection{Complexity of \cref{theo:reach_zon}}
First, we determine the complexity of \cref{theo:reach_zon}. 
The complexity of multiplying $\bar{A}_i \in \mathbb{R}^{n_{y} \times n_{y}}$ with ${\mathcal{Y}(k-i)}$ is upperbounded by $\mathcal{O}(f_k f_n^3 p)$ since $\mathcal{Y}(k-i)$ can have up to $(p+k_h) {on_{\tilde{u}}}$ generators. As this multiplication has to be done $p$ times at each time step $k$, the overall complexity is $\mathcal{O}(f_k^2 f_n^3 p^3)$.
The multiplication of $\bar{B}_i$ with ${\tilde{\mathcal{U}}(k-i)}$ has the overall complexity $\mathcal{O}(f_k f_n^3 p^2)$ as this multiplication is performed $(p+1)$ times at each time step and ${\tilde{\mathcal{U}}(k-i)}$ has ${on_{\tilde{u}}}$ generators.
Since we know the positions of dependent generators in the labeled generator matrices, we do not have to do any sorting for exact addition.
Summing $(2p+1)$ generator matrices with a maximum of $(p+k_h){on_{\tilde{u}}}$  generators results in the complexity $\mathcal{O}(f_k^2 f_n^2 p^3)$ for all time steps.  

From the partial complexities presented in this section, we conclude that \cref{theo:reach_zon} scales quadratically w.r.t. $f_k$ and cubically w.r.t. $f_n$ and $p$, as shown in \cref{tab:complexity}.

\subsection{Complexity of \cref{theo:reach_gen}}
To obtain the overall complexity of \cref{theo:reach_gen}, we examine the complexity of each step of \cref{alg:reachRec}. 
In \cref{algline:Atilde}, we have to calculate $\tilde{A}(k_{\text{init}})=\bar{A}_\text{ext}^{k_{\text{init}}}$. When multiplying $\bar{A}_\text{ext}$ with another $pn_y \times pn_y$-matrix, we only have to multiply the last $n_{y}$ rows of $\bar{A}_\text{ext}$ with the other matrix -- the remaining rows of the product are identical to the last $(p-1)n_{y}$ rows of the other matrix due to the special structure of $\bar{A}_\text{ext}$. This leads to a complexity of $\mathcal{O}(f_n^3 p^2)$ for one multiplication. When starting the algorithm with $k_{\text{init}}=p$, the complexity of computing $\tilde{A}(k_{\text{init}})$ can be upperbounded by $\mathcal{O}(f_n^3 p^3)$. This complexity could be further reduced using a more sophisticated matrix multiplication algorithm \cite{alman2021mult}. 
The complexity of the additions and multiplications in \cref{algline:Btilde0,algline:Btildei} is $\mathcal{O}(f_n^3 p^2)$ where $\bar{A}_\text{ext}^j$, $j=0,...,p$, is obtained as byproduct when computing $\tilde{A}(p)$.
\cref{algline:ARMAXtvp_S} has the complexity $\mathcal{O}(f_n^3 p^2)$.
\cref{algline:ARMAXtvp_reach} has to be executed $k_h/p$ times and thus, its overall computational complexity is $\mathcal{O}(f_k f_n^3 p^2)$.
If $k_{\text{init}}=p$, \cref{algline:ARMAXtvp_paramB2p,algline:ARMAXtvp_paramB2i} have to be executed once at $k=2p$ where the complexity of \cref{eq:ARMAXtvp_paramB} is $\mathcal{O}(f_n^3  p^2)$ and the complexity of \cref{eq:ARMAXtvp_paramRecB3} is $\mathcal{O}(f_n^3 p^3)$.
As the set $\mathcal{S}$ can have up to $k_h{on_{\tilde{u}}}$ generators, the complexity of executing \cref{algline:ARMAXtvp_SRec} for all time steps is $\mathcal{O}(f_k^2 f_n^3 p^3)$.
Summing up, \cref{alg:reachRec} scales quadratically w.r.t. $f_k$ and cubically w.r.t. $f_n$ and $p$.

\begin{table}[t]
\vspace{0.2cm}
    \caption{Computational complexity with symbolic zonotopes.}
    \label{tab:complexity}
    \centering
    \begin{tabular}{c c c c}
       \toprule
       \textbf{Approach} & \multicolumn{3}{c}{\textbf{Time complexity w.r.t.}}\\
        &  $f_k$ & $f_n$ & $p$  \\
       \midrule 
        \cref{theo:reach_zon} & $\mathcal{O}(f_k^2)$ & $\mathcal{O}(f_n^3)$ & $\mathcal{O}(p^3)$\\ 
        \cref{theo:reach_gen}  & $\mathcal{O}(f_k^2)$ & $\mathcal{O}(f_n^3)$  & $\mathcal{O}(p^3)$ \\ 
        \cref{theo:reach_eff}  & $\mathcal{O}(f_k)$ & $\mathcal{O}(f_n^3)$  & $\mathcal{O}(p^3)$ \\ 
        \bottomrule
    \end{tabular}
    \vspace{-0.5cm}
\end{table}

\subsection{Complexity of \cref{theo:reach_eff}}
Analogously to \cref{theo:reach_gen}, the complexities of \cref{algline:A,algline:B} in \cref{alg:approx} are $\mathcal{O}(f_n^3 p^3)$ and $\mathcal{O}(f_n^3 p^3 )$, respectively.
Initializing the sets $\mathcal{S}_c$, $\mathcal{T}_{c1}$ and $\mathcal{T}_{c2}$ in \cref{algline:Sc,algline:Tc1,algline:Tc2} has the complexity $\mathcal{O}(f_n^3 p^2)$.
The complexity of computing $\tilde{y}_v$ is $\mathcal{O}(f_n^2 p^2)$ for \cref{algline:sv} and $\mathcal{O}(f_k f_n^2 p^2)$ for evaluating \cref{algline:yv,algline:sv_rec} $k_h/p$ times.
By only applying linear transformations to the sets $\mathcal{T}_{c1}$ and $\mathcal{T}_{c2}$, the number of generators of these sets stays constant. Since $\mathcal{T}_{c1}$ and $\mathcal{T}_{c2}$ are initialized with $p{on_{\tilde{u}}}$ generators, the computational complexity of evaluating \cref{algline:Tc1_rec,algline:Tc2_rec} for all time steps is $\mathcal{O}(f_k f_n^3 p^2)$.
The sets $\mathcal{S}_c$, $\tilde{\mathcal{Y}}_c$ and $\tilde{\mathcal{Y}}_{(k:k_+)}$, on the other hand, are only modified via Minkowski sum in \cref{algline:Yc,algline:Ysplit,algline:Sc_rec}. As the computational complexity of Minkowski sum is independent of the number of generators of the summands, the overall complexity of these lines is $\mathcal{O}(f_k f_n p)$. 
Thus, \cref{alg:approx} scales linearly with $f_k$ and cubically w.r.t. $f_{n}$ and $p$.

%%%%%%%%%%%%%%%%%%%%%%%%%%%%%%%%%%%%%%%%%%%%%%%%%%%%%%%%%%%%%%%%%%%%%%%%%%%%%%%%
\section{NUMERICAL EXPERIMENTS} \label{sec:results}
In this section, we verify the proposed approaches with numerical experiments using symbolic zonotopes as set representation.
All computations are carried out in MATLAB on an i9-12900HK processor (2.5GHz) with 64GB memory.
The code to reproduce the results will be made publicly available with the next release of the toolbox CORA \cite{althoff2015introduction}.

\subsection{Accuracy} \label{sec:results_acc}

First, we consider a simple discrete-time SS model describing the motion of a pedestrian with
 \begin{align*}
 A = \begin{bmatrix}1 & 0 & 0.01 & 0 \\
0 & 1 & 0 & 0.01\\
0 & 0 & 1 & 0 \\
0 & 0 & 0 & 1 
\end{bmatrix},& \quad 
B = \begin{bmatrix}5\cdot 10^{-5} & 0\\
0 & 5\cdot 10^{-5}\\ 
0.01 & 0\\ 
0 & 0.01
\end{bmatrix},\\
C = \begin{bmatrix} 1 & 0 & 0 & 0 \\
0 & 1 & 0 & 0
\end{bmatrix},& \quad
D = \begin{bmatrix} 0 &  0 \\
0 &  0\end{bmatrix}.
\end{align*}
Using \cref{prop:transfSS_ARMAX} and the transformation matrix
\begin{align*}
M = \begin{bmatrix}-2 & 0 \\
0 & -2\\
-100 & 0\\
0 & -100\end{bmatrix},
\end{align*}
we can compute the parameters $\bar{A}_{i}$ and $\bar{B}_{i}$ of an equivalent ARMAX model with order $p=2$. 

To analyze the accuracy of the computed reachable sets, we compare the results of \cref{theo:reach_zon} (termed \emph{ARMAX}), which are identical to the results of \cref{theo:reach_gen} and \cref{theo:reach_eff}, with the traditional SS reachability formula from \cref{eq:SS_reach} (termed \emph{SS}).
To visualize the dependency problem, we also evaluate \cref{eq:ARMAXzon_reach} while replacing exact additions with Minkowski sums (termed \emph{ARMAX-DP}). 

As in many real-world situations, we assume that we do not know the initial system state, but the first $p$ measurements $\tilde{y}_\text{init}$ are given. Since the reachability formula of the SS model requires an estimate of the initial state set $\mathcal{X}(0)$, we have to estimate it from the measurements. This can be done by computing $\tilde{\mathcal{Y}}_{(0:p-1)}$ with \cref{eq:SS_reach} and solving for $\mathcal{X}(0)$:
\begin{align*}
    \mathcal{X}(0) &= \mathcal{O}_p^{\dag}
    \bigl(\tilde{\mathcal{Y}}_{(0:p-1)} \oplus(-\mathcal{H}(0) \times ... \times- {\mathcal{H}}(p-1))  \bigr)
\end{align*}
where $\mathcal{O}_p^{\dag}$ is the Moore-Penrose inverse of the observability matrix $\mathcal{O}_p=[C^T~(CA)^T~ ...~ (CA^{p-1})^T]^T$, the initial measurement sets are given as $\tilde{\mathcal{Y}}_{(0:p-1)}=\tilde{y}_\text{init}$, and
\begin{align*}
    \mathcal{H}(k)&=D \mathcal{U}(k) \oplus \mathcal{V}(k) \\ &\quad \oplus \bigoplus_{i=1}^{k} {C}{A}^{i-1} \bigl( {B}\mathcal{U}(k-i) \oplus \mathcal{W}(k-i)\bigr).
\end{align*}
Using the estimated $\mathcal{X}(0)$, the reachable sets $\mathcal{Y}(k)$, $k\geq p$, of the SS model can be computed with \cref{eq:SS_reach}. 

\begin{figure}[t]
\vspace{0.2cm}
\begin{subfigure}{.97\linewidth}
        \centering        \includegraphics[width=1\textwidth, keepaspectratio]{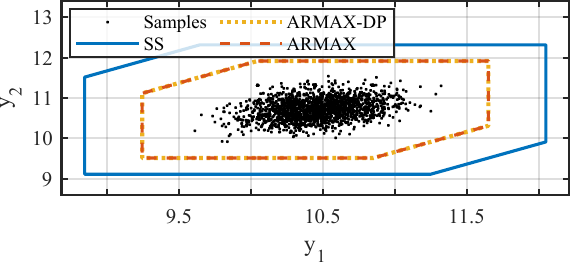}
        \vspace{-0.5cm}
        \caption{Reachable set at time step $k=p$.}
        \label{fig:reachresults0}
\end{subfigure}
\vspace{0.5cm}

\begin{subfigure}{.97\linewidth}
        \centering        \includegraphics[width=1\textwidth, keepaspectratio]{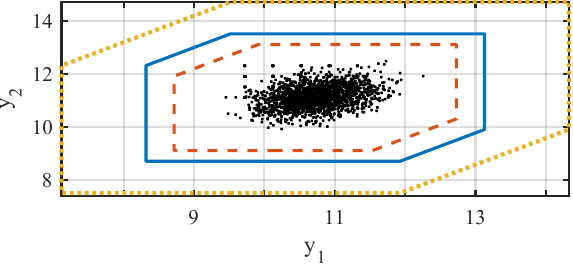}
        \vspace{-0.5cm}
        \caption{Reachable set at time step $k=p+1$.}
        \label{fig:reachresults1}
\end{subfigure}
\vspace{0.5cm}

\begin{subfigure}{.97\linewidth}
        \centering        \includegraphics[width=1\textwidth, keepaspectratio]{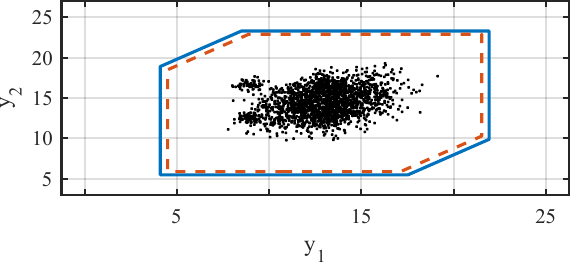}
        \vspace{-0.5cm}
        \caption{Reachable set at time step $k=p+9$.}
        \label{fig:reachresults6}
\end{subfigure}%
\vspace{0.1cm}
\caption{Reachability analysis of the pedestrian model.}
\label{fig:reachresults}
\vspace{-0.2cm}
\end{figure}

To investigate the tightness of the reachable sets, we simulate 2000 sample trajectories using \cref{eq:ARMAXtvp} with random disturbances $v(\cdot)\in \mathcal{V}$ and $w(\cdot)\in \mathcal{W}$ and a given input trajectory $u(\cdot)$.
The computed reachable sets and the sample trajectories are displayed in \cref{fig:reachresults}.
From \cref{fig:reachresults0}, we can see that the estimated reachable set of \emph{ARMAX-DP} and \emph{ARMAX} is identical for $k=p$, but over time, the predictions of \emph{ARMAX-DP} blow up due to the negligence of dependencies.
\emph{SS} starts with a marginally larger reachable set since information regarding the unknown disturbances $w(i)$, $i<p-1$, which is encoded in the measurements, is lost in the estimation of $\mathcal{X}_0$. 

\subsection{Scalability}
To investigate the scalability of the proposed approach, we measure the computation times while increasing the complexity of the reachability problem using random system matrices. In this regard, we study the consequences of a) increasing the time horizon by the factor $f_k$, b) the system dimensions by the factor $f_n$, and c) the order $p$ of the ARMAX model.
The nominal input and output dimension are  $n_{y0}=2$, $n_{\tilde{u}0}=3$. 
Due to implementation details and the inefficient matrix concatenation in Matlab, the computation times do not precisely scale according to the calculated complexities.
However, as shown in \cref{tab:compTimes}, \cref{theo:reach_zon,theo:reach_gen,theo:reach_eff} can be used for high-dimensional problems. \cref{theo:reach_eff} requires the shortest computation times in almost all scenarios. 

\begin{table}[hbt]
\vspace{0.2cm}
    \caption{Computation times.}
    \label{tab:compTimes}
    \centering
    \begin{tabular}{c c  c c c c}
       \toprule
       \multicolumn{3}{c}{} & \multicolumn{3}{c}{\textbf{Median computation time [s]}} \\
       ${f_k}$ & $f_n$ & $p$ & {\cref{theo:reach_zon}}& {\cref{theo:reach_gen}} & {\cref{theo:reach_eff}} \\
       \cmidrule(lr){1-3}\cmidrule(lr){4-6}
        4 & 1 & 2 & 0.0008 & 0.0007 & 0.0005 \\
        4000 & 1 & 2 & 55.7 & 17.6 & 8.1 \\
        16000 & 1 & 2 & 283 & 150 & 80 \\
        4 & 500 & 2 & 3.8 & 5.2 & 3.7 \\
        4 & 2000 & 2 & 173 & 303 & 205 \\
        4 & 1 & 150 & 0.9 & 3.6 & 0.9 \\
        4 & 1 & 600 & 291 & 216 & 29 \\
        \bottomrule
    \end{tabular}    
\end{table}

%%%%%%%%%%%%%%%%%%%%%%%%%%%%%%%%%%%%%%%%%%%%%%%%%%%%%%%%%%%%%%%%%%%%%%%%%%%%%%%%
\section{CONCLUSION} \label{sec:conclusion}

We presented three approaches to compute the reachable set of ARMAX models. The first method is for dependency-preserving set representations, while the second method can be used with arbitrary set representations based on a reformulation of the model. The third method we proposed leads to a reduced computational complexity. 
As demonstrated by numerical experiments, the resulting reachable sets are tighter than the reachable sets of equivalent state space models when the initial system state is unknown.

Going forward, we plan to extend our methodology to use ARMAX models in reachset-conformant model identification. 
Furthermore, our approach can be extended to handle other types of input-output models, which will be an exciting avenue for future work.
Overall, this work paves the way for more accurate and efficient verification methods for complex, data-defined systems.

\addtolength{\textheight}{-3cm}   % This command serves to balance the column lengths

\bibliographystyle{ieeetr}
\bibliography{literature}

\end{document}